\documentclass[aps,prb,twocolumn,twoside,a4paper,10pt,amsmath,amssymb,showpacs]{revtex4-1}

\usepackage{mathrsfs}
\usepackage{graphicx}
\usepackage{CJK}

\def\hatd#1{\hat{#1}^\dagger}
\def\bra#1{\left\langle{#1}\right|}
\def\ket#1{\left|{#1}\right\rangle}
\def\braket#1#2{\left\langle{{#1}}\mathrel{\left|{\vphantom{{#1}{#2}}}\right.\kern-\nulldelimiterspace}{{#2}}\right\rangle}

\begin{document}


\title{Topological magnon bound-states in periodically modulated Heisenberg XXZ chains}

\author{Xizhou Qin$^{1}$}
\author{Feng Mei$^{1}$}
\author{Yongguan Ke$^{1,2}$}
\author{Li Zhang$^{1,2}$}
\author{Chaohong Lee$^{1,2,}$}
\altaffiliation{Corresponding author. Email: lichaoh2@mail.sysu.edu.cn}

\affiliation{$^{1}$TianQin Research Center \& School of Physics and Astronomy, Sun Yat-Sen University (Zhuhai Campus), Zhuhai 519082, China}
\affiliation{$^{2}$Key Laboratory of Optoelectronic Materials and Technologies, Sun Yat-Sen University (Guangzhou Campus), Guangzhou 510275, China}

\date{\today}

\begin{abstract}
Strongly interacting topological states in multi-particle quantum systems pose great challenges to both theory and experiment.
Recently, bound states of elementary spin waves (magnons) in quantum magnets have been experimentally observed in quantum Heisenberg chains comprising ultracold Bose atoms in optical lattices.
Here, we explore a strongly interacting topological state called topological magnon bound-state in the quantum Heisenberg chain under cotranslational symmetry.
We find that the cotranslational symmetry is the key to the definition of a topological invariant for multi-particle quantum states, which enables us to characterize the topological features of multi-magnon excitations.
We calculate energy spectra, density distributions, correlations and Chern numbers of the two-magnon bound-states and show the existence of topological protected edge bound-states.
Our study not only opens a new prospect to pursue topological magnon bound-states, but also gives insights into the characterization and understanding of strongly interacting topological states.
\end{abstract}

\maketitle

\section{Introduction\label{Sec1}}

The study of magnons in quantum spin models can provide fundamental insights into collective excitations and low-energy properties of quantum magnetic systems.
The quantum Heisenberg chain is one of the typical models of interacting spins.
The well-known collective excitations around the ground state of a ferromagnetic Heisenberg chain appear as quasiparticle excitations, the so-called magnons~\cite{Bethe1931}.
Due to the ferromagnetic interaction, in addition to continuous scattering states, magnon bound-states do appear when the interaction is strong enough~\cite{Bethe1931, Wortis1963, Hanus1963}.
Recently, magnon excitations and their bound-states have been observed in the quantum Heisenberg chains comprising bosonic atoms in optical lattices~\cite{BlochNatPhys2013, BlochNature2013}.

On the other hand, since the discovery of topological insulators, the studies of topological matters have attracted extensive interests in recent years~\cite{KaneRev2010, QiRev2011, MooreNature2010}.
Up to now, theoretical understanding of noninteracting topological phases has been established completely~\cite{Schnyder2008, Kitaev2009}.
In the past few years, the topological feature of single-magnon excitations has been studied in both theory and experiment~\cite{MH2010, ExpMH2010, Li2013, Pereiro2014, Sachdev2014, Ong2015, Lee2015}.
Beyond the scenario for single-magnon topological excitations, it is of great importance to clarify whether topological bound-states can emerge in a strongly interacting multi-magnon system.
In general, strongly interacting topological states pose much difficulties to both theory~\cite{FTI2015} and experiment~\cite{Grusdt2015}.
The characterization of strongly interacting topological states is quite different from that of the weakly interacting counterparts~\cite{Chen2012, Wang2014}.
In particular, due to the strong inter-particle correlations, it is not easy to define a topological invariant and to clarify the interplay between topological features and inter-particle interactions.

In this article, we pursue a new direction to explore the topological magnon bound-states in a quantum Heisenberg XXZ chain.
We find that it is possible to define a topological invariant if the system presents a \textit{cotranslational symmetry}: collective translational invariance of a multi-particle system.
We demonstrate that our topological invariant can be used to characterize the topological feature of multi-magnon bound-states.

Compared with other topological invariants, our definition directly associates with the intrinsic symmetry in the considered system.
The famous Thouless-Kohmoto-Nightingale-Nijs (TKNN) invariant (i.e. the first Chern number)~\cite{TKNN}, which is an integral of the Berry curvature of Bloch states over the first Brillouin zone, strongly depends on the Bloch states, and thus applies only to noninteracting systems.
A well-known generalization of TKNN invariant for interacting systems is by using the so-called twisted boundary condition~\cite{Niu1985}.
Another way to defining topological invariant for interacting systems is the Green-function method~\cite{WangPRL2010, WangPRX2012}.
Differently, our topological invariant is directly defined based upon the cotranslational symmetry.
With our topological invariant, we study the two-magnon excitations in a periodically modulated Heisenberg XXZ chain and explore the emergence of topological magnon bound-states.
To give insights to the topological origin of these magnon bound-states, we derive an effective Hamiltonian by using many-body degenerate perturbation theory.
Moreover, it is possible to experimentally examine our predictions via quantum spin chains comprising ultracold atoms~\cite{BlochNatPhys2013, BlochNature2013} or ions~\cite{iontrapEXP1, iontrapEXP2, Grass2015}.

This article is organized as follows.
This section introduces the background and motivation.
In Sec.~\ref{Sec2}, we describe our model and discuss its basic properties.
In Sec.~\ref{Sec3}, we introduce the cotranslational symmetry in interacting multi-particle systems and define a topological invariant associates with this symmetry.
In Sec.~\ref{Sec4}, we study two-magnon excitations in a periodically modulated Heisenberg chain.
We give the two-magnon excitation spectrum, calculate the defined Chern numbers and reveal butterfly-like spectrum of magnon bound-states.
In order to clarify the topological origin of magnon bound-states in our model, we derive an effective single-particle Hamiltonian by using the degenerate perturbation theory in Sec.~\ref{Sec5}.
Finally, we discuss the experimental possibilities and summarize our results in Sec.~\ref{Sec6}.

\section{Model\label{Sec2}}

We consider a Heisenberg XXZ chain described by the following Hamiltonian,
\begin{eqnarray}\label{Eq_XXZ_Ham}
  \hat{H}&=&-\sum_{l}\left[J\big(\hat{S}_l^+\hat{S}_{l+1}^-+\hat{S}_l^-\hat{S}_{l+1}^+\big)
  +\Delta\hat{S}_l^z\hat{S}_{l+1}^z\right] \nonumber \\
  &&+\sum_{l}B_l\hat{S}_l^z+B_0\sum_{l}\hat{S}_l^z,
\end{eqnarray}
with the spin-$\frac{1}{2}$ operators $(\hat{S}_l^x, \hat{S}_l^y, \hat{S}_l^z)$ and $\hat{S}^\pm_l=\hat{S}^x_l\pm{i}\hat{S}^y_l$ for the $l$-th lattice site.
For simplicity, we use the units of the reduced Planck constant $\hbar=1$.
Here, $J$ and $\Delta$ are the transverse and longitudinal spin-exchange couplings, respectively.
In addition to the uniform field $B_0$, the magnetic fields include periodic modulations $B_l=\lambda\cos(2\pi\beta l+\delta)$ characterized by the modulation amplitude $\lambda$, the periodic parameter $\delta$ and the rational number $\beta=p/q$ (in which $p$ and $q$ are coprime integers).
Similar to the modulated single-particle systems~\cite{Lang2012}, the periodic parameter $\delta$ may act as an additional dimension for constructing the torus and defining the topological invariant.
However, differently, the single-particle quasi-momenta for interacting multi-particle systems are no longer good quantum numbers. This difference imposes a great challenge to define topological invariants for interacting multi-particle systems.

For convenience, we concentrate on studying magnon excitations over the fully magnetized state $\ket{\downarrow\downarrow\dots\downarrow}$ with all spins pointing downwards.
For a positive and sufficiently large $B_0$, the ground state is the fully magnetized state $\ket{\downarrow\downarrow\dots\downarrow}$.
The excited states over $\ket{\downarrow\downarrow\dots\downarrow}$ can be obtained by flipping spins.
Regarding the ground state as a vacuum state, the excited states can be viewed as a gas of quasi-particles called magnons.
Thus the transverse and longitudinal spin-exchange couplings respectively represent the hopping and interaction between magnons, and the magnetic fields $(B_0, B_l)$ describe the on-site energy of magnons.
Since the Hamiltonian~\eqref{Eq_XXZ_Ham} respects the $U(1)$ symmetry under rotation transformations, $\hat{S}_l^\pm\rightarrow e^{\pm i\theta}\hat{S}_l^\pm$, the total spin along $z$-direction $\hat{S}^z=\sum_l\hat{S}_l^z$ is conserved.
Therefore, the total magnon number (the number of spins that point upwards) $\hat{N}=\sum_l\hat{n}_l$ is also conserved, where $\hat{n}_l=\hat{S}_l^z+1/2$ is the magnon number for the $l$-th lattice site.

\section{Topological invariant under cotranslational symmetry\label{Sec3}}

The quantum Heisenberg chain~\eqref{Eq_XXZ_Ham} actually respects a cotranslational symmetry associated with the cotranslational operator.
Under periodic boundary condition (BC), the center-of-mass (c.o.m) quasi-momentum associated with the cotranslational operator is a good quantum number.
Based on the above c.o.m quasi-momentum and the periodic parameter of the modulation field, we successfully construct an artificial torus and then define a Chern topological invariant.

In order to define the cotranslational symmetry, we introduce the $N$-magnon wavefunction.
For a $L_t$-lattice system under the periodic BC, the $N$-magnon Hilbert space can be spanned by the basis,
$\mathcal{B}^{(N)}=\big\{\ket{l_1l_2\ldots l_N}=\hat{S}^+_{l_1}\hat{S}^+_{l_2}\ldots\hat{S}^+_{l_N}\ket{\mathbf{0}}\big\}$, in which $\ket{\mathbf{0}}=\ket{\downarrow\downarrow\dots\downarrow}$ and $1\le l_1<l_2<\ldots<l_N\le L_t$.
An arbitrary $N$-magnon state can be expanded as,
$\ket{\Psi}=\sum_{l_1<l_2<\ldots<l_N}\psi(l_1,l_2,\ldots,l_N)\ket{l_1l_2\ldots l_N}$,
where the $N$-magnon wavefunction
$\psi(l_1,l_2,\ldots,l_N)= \bra{\mathbf{0}}\hat{S}^-_{l_N}\ldots\hat{S}^-_{l_2}\hat{S}^-_{l_1}\ket{\Psi}$.
According to Hamiltonian (1), the $N$-magnon wavefunction obeys,
\begin{eqnarray}\label{Eq_eigen}
  H&&\psi(l_1,l_2,\ldots,l_N) \nonumber \\
  =&&-J\sum_{j=1}^{N}\sum_{\alpha=\pm 1}\psi(l_1,\ldots,l_{j-1},l_j+\alpha,l_{j+1},\ldots,l_N) \nonumber \\
  &&+V_{l_1l_2\ldots l_N}\psi(l_1,l_2,\ldots,l_N),
\end{eqnarray}
with
\begin{equation}
  V_{l_1l_2\ldots l_N}=-\Delta\sum_{j=1}^{N-1}\delta_{l_j+1,l_{j+1}}
  -\Delta\delta_{l_1,1}\delta_{l_NL_t}+\sum_{j=1}^{N}B_{l_j},
\end{equation}
where a constant energy shift has been removed.

In the $N$-magnon sector, the cotranslational operator $T_q(\tau)$ can be defined as
\begin{equation}\label{Eq_Cotrans_Op}
  T_q(\tau)\psi(l_1,\ldots,l_N)=\psi(l_1+\tau q,\ldots,l_N+\tau q),
\end{equation}
where $\tau$ is an integer.
Here we assume the lattice size $L_t$ is commensurate with the periodic modulation (i.e. $L_t=qs$ with $s$ an integer).
As $T_q(\tau)T_q(\tau')=T_q(\tau')T_q(\tau)=T_q(\tau+\tau')$ and $[T_q(\tau)]^{-1}=T_q(-\tau)$,
an Abelian group can be constructed.
We call this Abelian group as the cotranslation group.
Since $V_{l_1+q,l_2+q,\ldots,l_N+q}=V_{l_1l_2\ldots l_N}$, one can easily demonstrate that $T_q(\tau)H\psi(l_1,\ldots,l_N)=HT_q(\tau)\psi(l_1,\ldots,l_N)$ holds for an arbitrary $N$-magnon wavefunction $\psi(l_1,\ldots,l_N)$.
Therefore the Hamiltonian commutes with all cotranslation operators, that is, it is invariant under the cotranslation group
\begin{equation}
  [{T}_q(\tau)]^{-1}{H}{T}_q(\tau)={H},
\end{equation}
which represents a cotranslational symmetry.

Under the cotranslational symmetry, the Hamiltonian $H$ and the cotranslational operator $T_q(\tau)$ have common eigenstates.
The common eigenstates obey
\begin{equation}\label{Eq_eigeneq_H_T}
  T_q(\tau)\psi(l_1,l_2,\ldots,l_N)=c_q(\tau)\psi(l_1,l_2,\ldots,l_N),
\end{equation}
where $c_q(\tau)$ are eigenvalues of $T_q(\tau)$.
It is easy to find $c_q(\tau_1)c_q(\tau_2)=c_q\left(\tau_1+\tau_2\right)$ and $[c_q(\tau)]^{-1}=c_q(-\tau)$,
thus the eigenvalues could be chosen as the exponential function $c_q(\tau)=e^{i k \tau q}$ with the parameter $k$~\cite{Kannappan2009}.
Moreover, under periodic BC, $k=2\pi\alpha/L_t$ (with $\alpha=1, 2, \cdots, L_t/q$) becomes a good quantum number.
From Eqs.~\eqref{Eq_Cotrans_Op} and \eqref{Eq_eigeneq_H_T}, we have
\begin{eqnarray}\label{Eq_Bloch_like_theorem}
  \psi(l_1+\tau q,\ldots,l_N+\tau q)
  &=&T_q(\tau)\psi(l_1,\ldots,l_N), \nonumber \\
  &=&e^{i k \tau q}\psi(l_1,\ldots,l_N),
\end{eqnarray}
which resembles the Bloch theorem for single-particle systems with translational symmetry.
Therefore, $k$ is the eigenvalue of the generator for the cotranslation group and functions as the corresponding c.o.m quasi-momentum.
By defining $\psi(l_1,\ldots,l_N)=e^{i\frac{k}{N}(l_1+\cdots+l_N)}\phi(l_1,\ldots,l_N)$, from Eq.~\eqref{Eq_Bloch_like_theorem},
we find that $\phi(l_1,\ldots,l_N)$ is invariant under the cotranslation group, i.e., $\phi(l_1+\tau q,\ldots,l_N+\tau q)=\phi(l_1,\ldots,l_N)$.
This means that $\psi(l_1,\ldots,l_N)$ resembles the Bloch function.

Based upon the cotranslational symmetry and the corresponding c.o.m quasi-momentum, we now show how to formulate the topological invariant for our interacting multi-magnon system.
It is important to note that the periodic parameter $\delta$ has been introduced in the modulation field.
Therefore, through combining the periodic parameter $\delta$ with the c.o.m quasi-momentum $k$, one can construct an artificial torus in the two-dimensional momentum space,
where $\delta$ acts as the quasi-momentum in the second dimension~\cite{Lang2012, Mei2012, Zhu2013, Mei2014}.
Based on this artificial torus, similar to a realistic two-dimensional periodic system~\cite{TKNN}, we can define the Chern number as
\begin{equation}\label{Eq_Chern_num}
  C_n=\frac{1}{2\pi}\int_{0}^{2\pi/q}\mathrm{d}k\int_{0}^{2\pi}\mathrm{d}\delta\ \mathcal{F}_n(k,\delta),
\end{equation}
where
$\mathcal{F}_n(k,\delta) =\mathrm{Im}\big(\braket{\partial_\delta\phi_n}{\partial_k\phi_n} -\braket{\partial_k\phi_n}{\partial_\delta\phi_n}\big)$
is the Berry curvature of the $n$-th band, with $\ket{\phi_n}=\ket{\phi_{n,k,\delta}}$ and
$\phi_{n,k,\delta}(l_1,\ldots,l_N) =\bra{\mathbf{0}}\hat{S}^-_{l_N} \ldots\hat{S}^-_{l_2}\hat{S}^-_{l_1} \ket{\phi_{n,k,\delta}}$.

\section{Topological two-magnon excitations\label{Sec4}}

\begin{figure*}[t]
  \includegraphics[width=2.0\columnwidth]{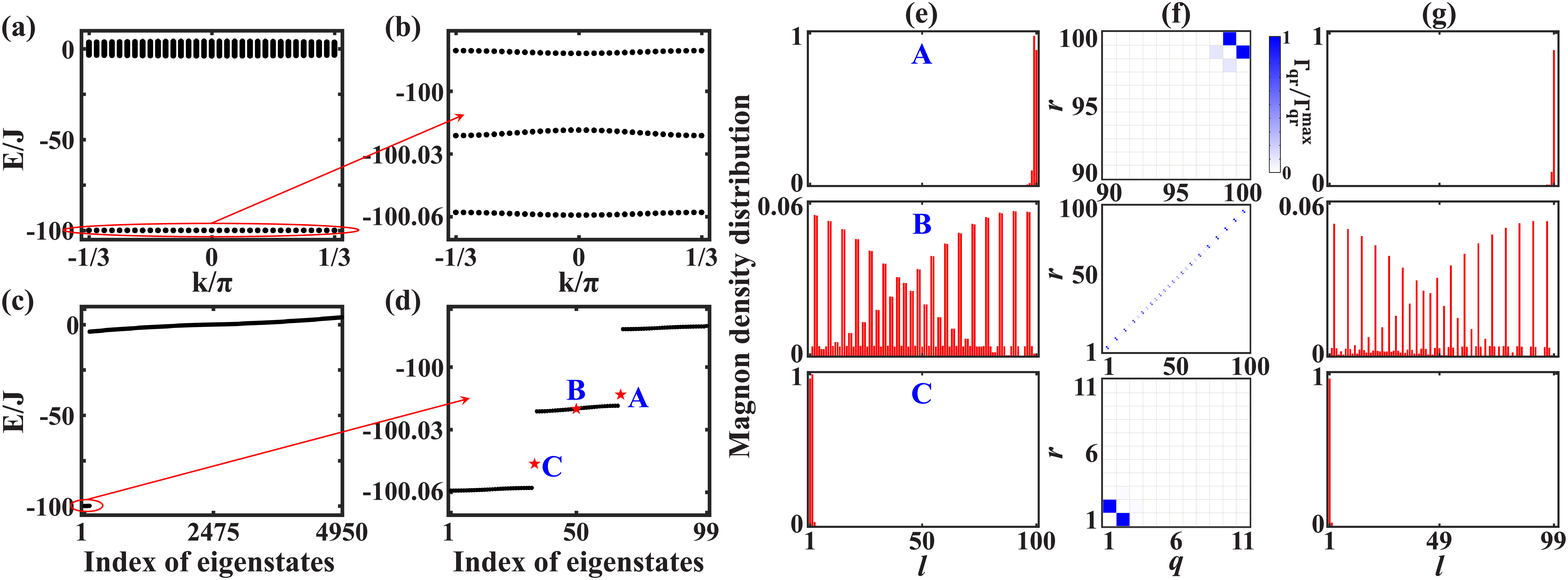}
  \caption{\label{Fig1}(Color online) Two-magnon excitation spectra.
  (a) Spectrum for periodic BC ($L_t=99$).
  (b) Magnon-bound-state spectrum for periodic BC ($L_t=99$).
  (c) Spectrum for open BC ($L_t=100$).
  (d) Magnon-bound-state spectrum for open BC ($L_t=100$).
  (e) Magnon density distribution $n_l=\bra{\psi}\hat{n}_l\ket{\psi}$ corresponding to the eigenstates of the red pentagrams (A - edge state localized at right, B - bulk state, C - edge state localized at left) in the spectrum (d).
  (f) Two-magnon correlations $\Gamma_{qr}$ corresponding to (e) (here only the region with non-zero correlations is presented).
  (g) Two-magnon minor diagonal correlations $\Gamma_{l,l+1}$ of (f).
  The other parameters are chosen as $\Delta/J=100$, $\lambda/J=0.04$, $\beta=1/3$ and $\delta=\pi/6$.}
\end{figure*}

Below we take two-magnon excitations as an example and show how topological bound-states emerge.
The two-magnon excitations in the field-free quantum Heisenberg chain have been well studied, such as the time-evolution dynamics can be understood as the quantum walks of two hard-core bosons~\cite{Qin2014}.
Under strong anisotropy ($\Delta\gg J>0$), the two-magnon spectrum includes two bands, in which the upper and lower bands respectively correspond to scattering and bound-states.
Below, we will study the two-magnon excitations in Hamiltonian~\eqref{Eq_XXZ_Ham} and demonstrate the existence of topological two-magnon bound-states via calculating energy spectra, density distributions, correlations and Chern numbers.

Given the two-magnon basis,
$\mathcal{B}^{(2)}=\big\{\ket{l_1l_2}=\hat{S}_{l_1}^+\hat{S}_{l_2}^+\ket{\mathbf{0}}\big\}$ with ($1\le l_1<l_2\le L_t$),
the two-magnon wavefunction can be written as $\ket{\Psi}=\sum_{l_1<l_2}\psi(l_1,l_2)\ket{l_1l_2}$.
Through Eq.~\eqref{Eq_eigen}, one can obtain the two-magnon eigenequation,
\begin{eqnarray}\label{Eq_Harper_2}
  H\psi(l_1,l_2)&=&-J\big[\psi(l_1+1,l_2)+\psi(l_1-1,l_2) \nonumber \\
  &&~~~~~+\psi(l_1,l_2+1)+\psi(l_1,l_2-1)\big] \nonumber \\
  &&+V_{l_1l_2}\psi(l_1,l_2)=E\psi(l_1,l_2),
\end{eqnarray}
where $V_{l_1l_2}=-\Delta(\delta_{l_1+1,l_2}+\delta_{l_1,1}\delta_{l_2,L_t})+(B_{l_1}+B_{l_2})$.
According to Eq.~\eqref{Eq_Bloch_like_theorem}, the two-magnon wavefunction reads as $\psi_{n,k}(l_1,l_2)=e^{\frac{i}{2}(l_1+l_2)k}\phi_{n,k}(l_1,l_2)$,
in which $\phi_{n,k}(l_1+q,l_2+q)=\phi_{n,k}(l_1,l_2)$.
Thus the above two-magnon eigenequation~\eqref{Eq_Harper_2} can be rewritten as
\begin{eqnarray}
  &&E_n\phi_{n,k}(l_1,l_2)=V_{l_1l_2}\phi_{n,k}(l_1,l_2)\nonumber\\
  &&-J\big[e^{\frac{i}{2}k}\phi_{n,k}(l_1+1,l_2) +e^{-\frac{i}{2}k}\phi_{n,k}(l_1-1,l_2)\nonumber \\
  &&~~~~~+e^{\frac{i}{2}k}\phi_{n,k}(l_1,l_2+1) +e^{-\frac{i}{2}k}\phi_{n,k}(l_1,l_2-1)\big].
\end{eqnarray}
Since $\phi_{n,k}(l_1+q,l_2+q)=\phi_{n,k}(l_1,l_2)$, the above eigenvalue problem reduces to the problem of diagonalizing a $\frac{q(L_t-1)}{2}\times\frac{q(L_t-1)}{2}$ matrix for each $k$.

We show the two-magnon excitation spectrum in Fig.~\ref{Fig1}.
Under periodic BC, there are two separated bands: the upper continuum band and the lower bound-state band, see Fig.~\ref{Fig1} (a).
The bound-state bulk band includes three subbands due to $\beta=1/3$ (it includes $q$ subbands if $\beta=p/q$), see Fig.~\ref{Fig1} (b).
Since the subbands are well separated from each other with finite gaps, we can successfully calculate their Chern numbers according to the definition~\eqref{Eq_Chern_num}.
In Tab.~\ref{Tab-TwoMagnonChernNumber}, we give the Chern numbers for $\beta=p/q$ for all odd $q$ with $3\le q\le 9$.
As the transformation $\beta \rightarrow (1-\beta)$ results to $\lambda\cos(2\pi\beta l+\delta) \rightarrow \lambda\cos(2\pi\beta l-\delta)$, one can find that the spectrum structure remains unchanged after the transformation, thus the systems with $\beta$ and $(1-\beta)$ have the same Chern numbers.
This means that we only need calculate the Chern numbers for $0<\beta<1/2$.

\begin{table}
  \caption{\label{Tab-TwoMagnonChernNumber} The Chern numbers for $q$ magnon-bound-state subbands. The parameters are chosen as $\beta=p/q$ ($3\le q\le 9$ and $q$ are odd numbers), $\Delta/J=100$, $\lambda/J=0.04$ and $L_t=qs>900$.}
  \begin{ruledtabular}
    \begin{tabular}{crrrrrrrrr}
      $p/q$ & $1/3$ & $1/5$ & $2/5$ & $1/7$ & $2/7$ & $3/7$ & $1/9$ & $2/9$ & $4/9$ \\
      \hline
            &       &       &       &       &       &       &  $1$  & $-4$  & $-2$  \\
            &       &       &       &  $1$  & $-3$  & $-2$  &  $1$  &  $5$  & $-2$  \\
            &       &  $1$  & $-2$  &  $1$  &  $4$  &  $5$  &  $1$  & $-4$  &  $7$  \\
            &  $1$  &  $1$  &  $3$  &  $1$  & $-3$  & $-2$  &  $1$  &  $5$  & $-2$  \\
      $C_n$ & $-2$  & $-4$  & $-2$  & $-6$  &  $4$  & $-2$  & $-8$  & $-4$  & $-2$  \\
            &  $1$  &  $1$  &  $3$  &  $1$  & $-3$  & $-2$  &  $1$  &  $5$  & $-2$  \\
            &       &  $1$  & $-2$  &  $1$  &  $4$  &  $5$  &  $1$  & $-4$  &  $7$  \\
            &       &       &       &  $1$  & $-3$  & $-2$  &  $1$  &  $5$  & $-2$  \\
            &       &       &       &       &       &       &  $1$  & $-4$  & $-2$  \\
    \end{tabular}
  \end{ruledtabular}
\end{table}

\begin{figure}[t]
  \includegraphics[width=1.0\columnwidth]{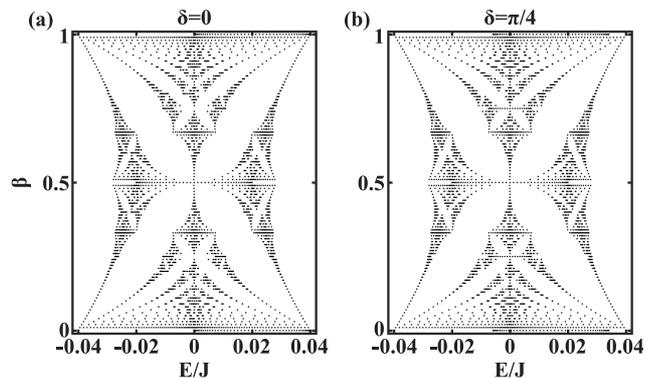}
  \caption{\label{Fig2} The butterfly-like spectrum for the two-magnon bound-states.
  We diagonalize the Hamiltonian~\eqref{Eq_XXZ_Ham} with $\Delta/J=100$, $L_t=100$ and different values of $\delta$:
  (a) $\delta=0$ and (b) $\delta=\pi/4$.
  The values of $\lambda$ are set as $\lambda/J=J[\Delta\cos{(\pi\beta)}]^{-1}$ for $\beta\ne1/2$, and $\lambda=0$ for $\beta=1/2$.
  The energies $E/J$ are added with a constant $\Delta/J+2J/\Delta$.}
\end{figure}

Our calculation shows that the three bound-state subbands for $\beta=1/3$ shown in Fig.~\ref{Fig1}b have the Chern numbers $C_n=(1,-2,1)$, respectively.
The nontrivial Chern numbers indicate that the magnon bound-states in these three subbands are nontrivial topological states.
Meanwhile, one can find that the Chern number distributions of these subbands resemble the ones of the Harper-Hofstadter model~\cite{Harper1955, Hofstadter1976}, which has been used to describe the integer quantum Hall effect~\cite{TKNN, Hatsugai1993a, Hatsugai1993b}.
This feature is quite remarkable because our system is of strong inter-particle interactions and the Harper-Hofstadter model is interaction-free.
The identical topological invariants of the two models means that they share the same topological origins.
This point is further supported by the butterfly-like spectrum for the two-magnon states, see Fig.~\ref{Fig2}.
Interestingly, the butterfly-like spectrum also resembles the one for the Harper-Hofstadter model.

\begin{figure}[t]
  \includegraphics[width=1.0\columnwidth]{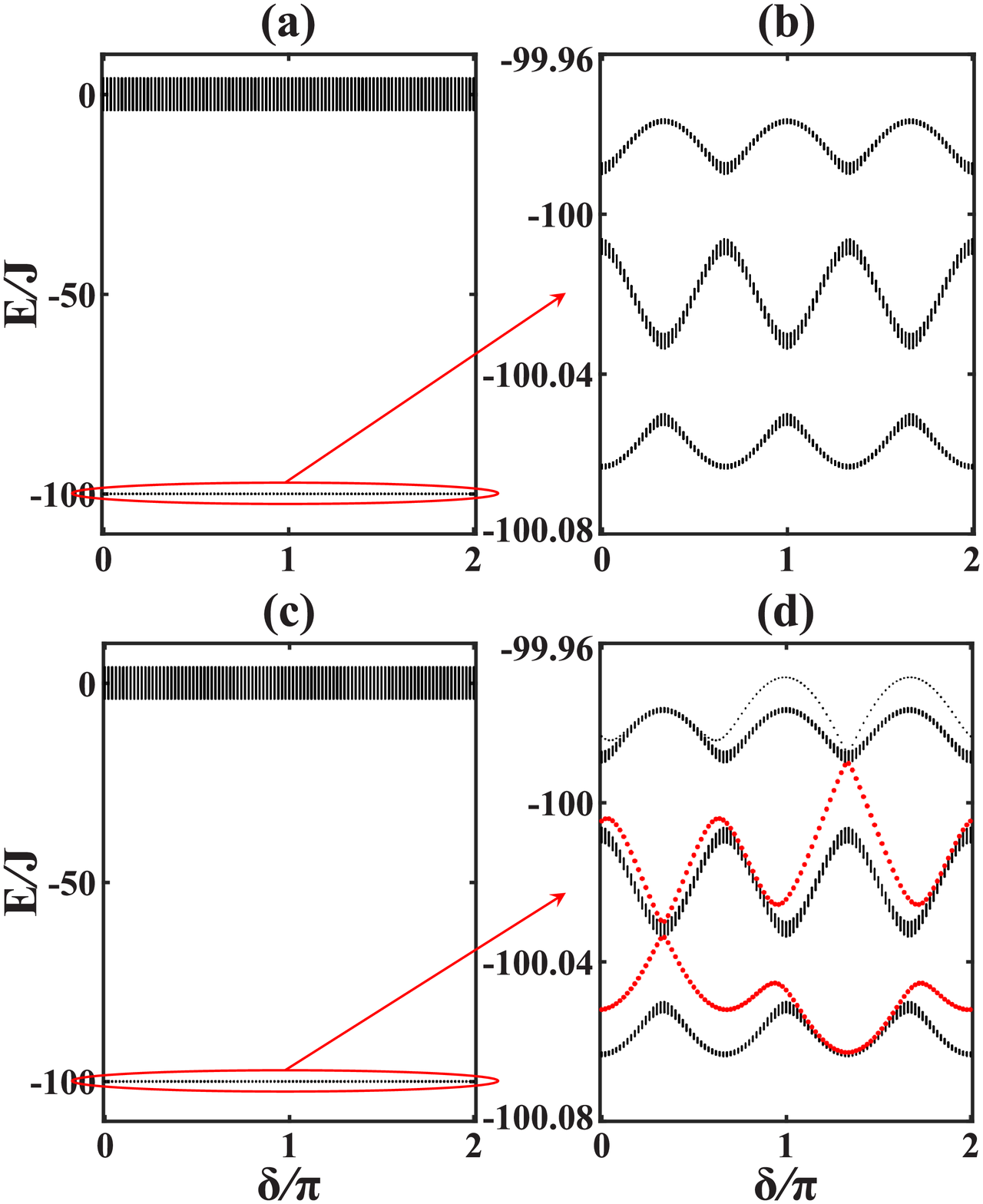}
  \caption{\label{Fig3}(Color online) The two-magnon spectrum versus the periodic parameter $\delta$.
  (a-b) periodic BC ($L_t=99$); (c-d) open BC ($L_t=100$).
  (b) and (d) are the enlarged magnon-bound-state spectra in (a) and (c), respectively.
  The other parameters are chosen as $\Delta/J=100$, $\lambda/J=0.04$, and $\beta=1/3$.}
\end{figure}

In addition to the nontrivial Chern numbers, another hallmark of topological phases is the appearance of edge states in the system under open BC.
Due to the open BC, the c.o.m quasi-momentum $k$ is no longer a good quantum number, see Fig.~\ref{Fig1} (c).
Besides the extended bound-states, the edge bound-states appear in the energy gaps, see the pentagrams A and C in Fig.~\ref{Fig1} (d).
To illustrate these bound-states are edge states, we plot their density distribution in Fig.~\ref{Fig1} (e).
The density distribution clearly shows that these bound-states do localize on the left or right.
In Fig.~\ref{Fig1} (f), we calculate the two-magnon spin correlations $\Gamma_{qr}=\bra{\psi}\hat{S}_q^+\hat{S}_r^+\hat{S}_r^-\hat{S}_q^-\ket{\psi}$ for localized edge and extended bulk bound-states.
The left and right edge bound-states manifest as the correlation peaks on lower left and upper right quarters along the two minor diagonal lines, respectively.
In contrast, the bulk bound-states extensively distribute along the whole minor diagonal lines.
In Fig.~\ref{Fig1} (g), we show the two-magnon correlations along the minor diagonal line $\Gamma_{l,l+1}$.
Moreover, when the modulation parameter $\delta$ changes from $0$ to $2\pi$, the two-magnon spectrum varies continuously and periodically with a period $2\pi/q$.
In Fig.~\ref{Fig3}, we show how the two-magnon spectrum changes with $\delta$.
According to the bulk-edge correspondence, the topological bulk state under periodic BC corresponds to the edge state under open BC.
We do find one edge state traversing in the energy gap for the system under open BC.

It is important to clarify the role of interaction in the emerging of topological magnon bound-states.
Under strong interaction, the energy spectrum will be separated into two bands: the upper band for continuum states and the lower band for bound-states, see Fig.~\ref{Fig1} (a).
Our calculation shows that the topological states lying in the lower band are bound-states.
This means that the topological states in our interacting system are indeed interaction-assisted.
Moreover, the interaction-assisted topological states are not limited to two-magnon excitations, but also could be generalized to multi-magnon excitations, which provide extra robustness for these topological bound-states.
In contrast, most topological states found in condensed matter physics will be destroyed when inter-particle interaction has been introduced.
Thus, our study adds a new member for interaction-enabled topological states and also sheds light on understanding interacting topological states.

\section{Effective single-particle Hamiltonian\label{Sec5}}

To give an effective picture for the topological magnon bound-states and reveal the resemblance between our model and the Harper-Hofstadter model~\cite{Harper1955, Hofstadter1976}, we derive an effective Hamiltonian by employing the many-body perturbation theory.
Under strong anisotropy ($|J/\Delta|\ll1$ and $|\lambda/\Delta|\ll1$), which corresponds to strong interactions between magnons, one can treat the sum of the hopping term and the modulation term,
\begin{equation}\label{Eq_hopping_external}
  \hat{H}_1=-J\sum_l\big(\hat{S}_l^+\hat{S}_{l+1}^-+\hat{S}_l^-\hat{S}_{l+1}^+\big)+\sum_{l}B_l\hat{S}_l^z,
\end{equation}
as a perturbation to the sum of the interaction term and the uniform field term,
\begin{equation}\label{Eq_interaction}
  \hat{H}_0=-\Delta\sum_{l}\hat{S}_l^z\hat{S}_{l+1}^z+B_0\sum_{l}\hat{S}_l^z.
\end{equation}
Within a many-body perturbation theory up to second order (see Appendix~\ref{SecAppB}), the effective Hamiltonian reads as
\begin{equation}\label{Eq_eff}
  \hat{H}_\mathrm{eff}=-\frac{J^2}{\Delta}\sum_{m=1}^{L_t}(\hatd{b}_m\hat{b}_{m+1}+\mathrm{H.c.})
  +\sum_{m=1}^{L_t}\mu_m\hatd{b}_m\hat{b}_m,
\end{equation}
where $\mu_m=\lambda'\cos(2\pi\beta m+\delta')$ with the parameters $\lambda'=2\lambda\cos(\pi\beta)$ and $\delta'=\delta+\pi\beta$.
In our derivation, we have chosen the periodic BC and omitted a constant energy shift.

In the effective Hamiltonian, the two-magnon bound-states are described as a single-particle states, which are associated with the creation operator $\hatd{b}_m$.
Obviously, $\hatd{b}_m\ket{\mathbf{0}}$ is equivalent to $\hat{S}^+_m\hat{S}^+_{m+1}\ket{\downarrow\downarrow\ldots\downarrow}$.
This means that, $\hatd{b}_m$ simultaneously flips two neighbouring spins: one on the $m$-th lattice site and the other one on the ($m+1$)-th lattice site.
It is easy to find that the form of the effective Hamiltonian is same as the Harper-Hofstadter model~\cite{Harper1955, Hofstadter1976}.
Thus, the origin of topological magnon bound-states can be understood from the integer quantum Hall states in the Harper-Hofstadter model.
That is, the topological magnon bound-state can be looked as an analogue to integer quantum Hall state.
Therefore, the corresponding topological features could be explained by the Chern topological invariant and the edge states (see Appendix~\ref{SecAppB}).
The topological features given by the effective Hamiltonian are well consistent with the numerical ones without using the perturbation theory, which are presented in the previous section.

Since the effective Hamiltonian is derived in the context of the second-order perturbation theory, it is unclear whether the perturbation treatment retains the topological nature of the original system.
Through introducing an auxiliary Hamiltonian, which is a combination of the effective Hamiltonian and the original Hamiltonian, we demonstrate the topological equivalence between the effective system and the original system (see Appendix~\ref{SecAppC}).

\section{Summary and discussions\label{Sec6}}

In summary, we have explored the existence of topological magnon bound-states in strongly interacting quantum spin chains under cotranslational symmetry.
We find that the cotranslational symmetry allows us to define a topological invariant for the interacting multi-magnon system.
Through analyzing their topological features, we demonstrate that topological magnon bound-states do emerge and they are interaction-enabled.
By employing the many-body degenerate perturbation theory, we derive an effective model for the topological magnon bound-states which resembles the interaction-free integer quantum Hall model, thus the topological magnon bound-states could be understood as quantum Hall states of bounded magnons.
Our study not only provides a new prospect to explore topological magnon excitations, but also sheds lights on understanding strongly interacting topological states.

Based on the current state-of-art in well-controlled many-body quantum systems, it is possible to realize our quantum spin chain with ultracold atoms.
In recent years, ultracold atoms have provided a leading platform for simulate various many-body quantum models in condensed matter physics and other fields.
The Heisenberg XXZ chain has been experimentally realized with two-component bosonic atoms in a one-dimensional optical lattice~\cite{BlochNatPhys2013, BlochNature2013}, in which both single-magnon and magnon-bound-states have been experimentally observed.
To engineer the desired on-site modulation on the systems~\cite{BlochNatPhys2013, BlochNature2013}, one can apply an additional state-dependent optical lattice~\cite{Mandel2003}, the Hamiltonian reads as
\begin{eqnarray}
  \hat H_\mathrm{BH} &=& -\sum_{\left\langle{j,k}\right\rangle,\sigma=\uparrow,\downarrow}t_\sigma\hatd b_{j\sigma}\hat b_{k\sigma}+\sum_{j\sigma}\frac{U_{\sigma\sigma}}{2}\hat n_{j\sigma}(\hat n_{j\sigma}-\hat 1) \nonumber \\
  &+& U_{\uparrow\downarrow}\sum_j\hat n_{j\uparrow}\hat n_{j\downarrow}+\sum_{j\sigma}V_{j\sigma}\hat n_{j\sigma},
\end{eqnarray}
where the state-dependent optical lattice potential gives $V_{j\uparrow}=-V_{j\downarrow}=\lambda\cos(2\pi\beta j+\delta)$.
Using the experimental techniques in~\cite{BlochNatPhys2013, BlochNature2013}, one can prepare an atomic Mott insulator which obeys the spin chain model~\eqref{Eq_XXZ_Ham} with the parameters
\begin{equation}
  J=\frac{2t_\uparrow t_\downarrow}{U_{\uparrow\downarrow}},
  \ \ \Delta=\frac{4t_\uparrow^2}{U_{\uparrow\uparrow}}
  +\frac{4t_\downarrow^2}{U_{\downarrow\downarrow}}
  -\frac{2(t_\uparrow^2+t_\downarrow^2)}{U_{\uparrow\downarrow}},
\end{equation}
where the pseudospin operators can be defined as $\hat{S}^+_l=\hatd b_{l\uparrow}\hat b_{l\downarrow}$, $\hat{S}^-_l=\hatd b_{l\downarrow}\hat b_{l\uparrow}$ and $\hat{S}^z_l=(\hat n_{l\uparrow}-\hat n_{l\downarrow})/2$.
The interaction strength between magnons can be tuned by Feshbach resonance~\cite{Widera2004, Gross2010}.

It is also possible to realize our quantum spin chain and observe the topological magnon bound-states with ultracold trapped ions.
In last years, the rapidly developing techniques in engineering ultracold trapped ions have become sufficiently mature to simulate the magnetic phases and the quantum dynamics in Heisenberg spin chains~\cite{iontrapEXP1, iontrapEXP2, Grass2015}.

\textit{Note added}: During the finalization of the present version (see Ref.~\cite{Qin2016} for our previous version), we became aware of the recent studies on two-particle topological states in an interacting bosonic~\cite{DiLiberto2016} or fermionic~\cite{Marques2017} SSH chain.
In Ref.~\cite{DiLiberto2016}, the c.o.m momentum of the two interacting particle is used as a good quantum number for calculating the two-particle generalized Zak phase, which is very similar to the use of c.o.m momentum for calculating our defined Chern number.
In Ref.~\cite{Marques2017}, an effective Hamiltonian for two-hole states in strongly interacting systems are given, which connects the two-hole model to a noninteracting single-hole model.

\acknowledgments

We thank T. Gra{\ss} for providing us their recent work~\cite{Grass2015} on the experimental realization of one-dimensional spin systems with synthetic magnetic fluxes.
This work is supported by the National Basic Research Program of China (Grant No. 2012CB821305) and the National Natural Science Foundation of China (Grant No. 11374375, 11574405).

\appendix

\section{Calculation of the Chern number\label{SecAppA}}

Below we give the procedure of calculating the Chern number via our definition \eqref{Eq_Chern_num}.
Firstly, the 2D parameter space is discretized as $\{k_\alpha=2\pi\alpha/L_t, \delta_j=2\pi{j}/N_\delta\}$ with $\alpha=\{1,2,\ldots,L_t/q\}$ and $j=\{1,2,\dots,N_\delta\}$, thus the Chern number is given as
\begin{equation}
  C_n=\frac{1}{2\pi}\sum_{\alpha,j}\mathcal{S}_{\alpha,j}(\mathcal{F}_n),
\end{equation}
with
\begin{equation}
  \mathcal{S}_{\alpha,j}(\mathcal{F}_n)=\int_{k_\alpha}^{k_{\alpha+1}}\mathrm{d}k \int_{\delta_j}^{\delta_{j+1}}\mathrm{d}\delta\ \mathcal{F}_n(k,\delta),
\end{equation}
being the integral of the Berry curvature over the rectangular region, $R_{\alpha,j}=[k_\alpha,k_{\alpha+1}]\times[\delta_j,\delta_{j+1}]$.
According to Stokes' formula, the integral $\mathcal{S}_{\alpha,j}(\mathcal{F}_n)$ is equal to the line integral of the Berry connection, $(\mathcal{A}_k,\mathcal{A}_\delta) =i\bra{\phi_{k,\delta}}(\partial_k,\partial_\delta)\ket{\phi_{k,\delta}}$, along the boundary of $R_{\alpha,j}$ (denoted as $B_{\alpha,j}$), i.e., the Berry phase along $B_{\alpha,j}$.

The Berry phase can be given as the limit of the discrete geometric phase as $N_\delta\rightarrow\infty$ and $L_t\rightarrow\infty$~\cite{Resta2000}.
If the integral region $R_{\alpha,j}$ is sufficiently small, the Berry phase along $B_{\alpha,j}$ can be approximately replaced by the discrete geometric phase along the four vertices of the rectangular region $R_{\alpha,j}$~\cite{Fukui2005}, that is,
\begin{equation}
  \mathcal{S}_{\alpha,j}(\mathcal{F}_n)=-\mathrm{Im}\Big[\ln\big(U_{12}U_{23}U_{34}U_{41}\big)\Big].
\end{equation}
Here, $U_{jj'}=\braket{\phi_j}{\phi_{j'}}/\left|\braket{\phi_j}{\phi_{j'}}\right|$,
$\ket{\phi_1}=\ket{\phi_{k_{\alpha},\delta_{j}}}$,
$\ket{\phi_2}=\ket{\phi_{k_{\alpha+1},\delta_{j}}}$,
$\ket{\phi_3}=\ket{\phi_{k_{\alpha+1},\delta_{j+1}}}$,
$\ket{\phi_4}=\ket{\phi_{k_{\alpha},\delta_{j+1}}}$,
and the principal branch of the natural logarithm is specified in $(-\pi,\pi]$.

\section{Derivation of the effective Hamiltonian\label{SecAppB}}

Under strong anisotropy ($|J/\Delta|\ll1$ and $|\lambda/\Delta|\ll1$), the two magnons can form a magnon bound-state and thus can be recognized as a single quasiparticle.
In this section, we will use the 2nd order perturbation theory for many-body degenerate quantum systems~\cite{Takahashi1977} to derive an effective Hamiltonian for the two-magnon bound-states.

As $|J/\Delta|\ll1$ and $|\lambda/\Delta|\ll1$, separating the Hamiltonian~\eqref{Eq_XXZ_Ham} as $\hat{H}=\hat{H}_0+\hat{H}_1$, one can treat
\begin{equation}\label{Eq_hopping_external_S}
  \hat{H}_1=-J\sum_l\big(\hat{S}_l^+\hat{S}_{l+1}^-+\hat{S}_l^-\hat{S}_{l+1}^+\big) +\sum_{l}B_l\hat{S}_l^z.
\end{equation}
as a perturbation to
\begin{equation}
  \hat{H}_0=-\Delta\sum_{l}\hat{S}_l^z\hat{S}_{l+1}^z+B_0\sum_{l}\hat{S}_l^z.
\end{equation}
For clarity, we first assume the system~\eqref{Eq_XXZ_Ham} obeying the periodic BC.

To implement the perturbation treatment, we give the projection operator onto the subspace involved by the magnon bound-states and the projection operator onto the orthogonal component of the involved subspace.
The unperturbed term $\hat{H}_0$ has only two eigenvalues: (i) $E_0=-\Delta(\frac{1}{4}L_t-1)-B_0(\frac{1}{2}L_t-2)$ for the $L_t$-fold degenerated ground-states $\{\ket{G_l}=\ket{l,l+1}:1\le l\le L_t\}$
and (ii) $E_{l_1l_2}=-\Delta(\frac{1}{4}L_t-2)-B_0(\frac{1}{2}L_t-2)$ for excited eigenstates $\{\ket{E_{l_1l_2}}=\ket{l_1,l_2}:l_1+1<l_2,\ (l_1,l_2)\ne(1,L_t)\}$.
The two-magnon bound-states only involve the subspace $\mathcal{U}_0$ spanned by $L_t$ independent ground states $\ket{G_l}$.
The projection operator onto $\mathcal{U}_0$ reads as
\begin{equation}\label{Eq_Project_U0_S}
  \hat{P}_0=\sum_l\ket{G_l}\bra{G_l}.
\end{equation}
Introducing $\mathcal{V}_0$ as the orthogonal complement of $\mathcal{U}_0$, the projection operator onto $\mathcal{V}_0$ reads as
\begin{equation}
  \hat{S}=\sum_{E_{l_1l_2}\ne E_0}\frac{1}{E_0-E_{l_1l_2}}\ket{E_{l_1l_2}}\bra{E_{l_1l_2}}.
\end{equation}
Using the notations $\hat{P}_0$ and $\hat{S}$ and including the perturbations up to second order~\cite{Takahashi1977}, the effective Hamiltonian is given as,
\begin{equation}\label{Eq_H_eff_def_S}
  \hat{H}_\mathrm{eff}=\hat{h}_0+\hat{h}_1+\hat{h}_2
  =E_0\hat{P}_0+\hat{P}_0\hat{H}_1\hat{P}_0
  +\hat{P}_0\hat{H}_1\hat{S}\hat{H}_1\hat{P}_0.
\end{equation}
One can benefit from the fact that the transverse coupling term only contributes to $\hat{h}_2$ while the modulation term only contributes to $\hat{h}_1$.

Using Eqs.~\eqref{Eq_hopping_external_S}~and~\eqref{Eq_Project_U0_S}, the first-order perturbation reads as
\begin{equation}
  \hat{h}_1=\hat{P}_0\hat{H}_1\hat{P}_0=\sum_{mm'}
  \ket{G_m}\bra{G_m}\hat{H}_1\ket{G_{m'}}\bra{G_{m'}}.
\end{equation}
Since $\bra{G_m}\hat{S}_l^z\ket{G_{m'}}=\delta_{mm'}(\delta_{l,m}+\delta_{l,m+1}-\frac{1}{2})$,
we have $\bra{G_m}\hat{H}_1\ket{G_{m'}}=\delta_{mm'}(B_m+B_{m+1}-\frac{1}{2}\sum_lB_l)$
and so that
\begin{equation}\label{Eq_h1_S}
  \hat{h}_1=\sum_{m=1}^{L_t}\left(B_m+B_{m+1}-\tfrac{1}{2}\sum_lB_l\right)\ket{G_m}\bra{G_m}.
\end{equation}

Since $\bra{E_{l_1l_2}}\hat{S}^z_{l'}\ket{G_{m'}}=0$, the second-order perturbation $\hat{h}_2=\hat{P}_0\hat{H}_1\hat{S}\hat{H}_1\hat{P}_0$ reads as
\begin{eqnarray}
  \hat{h}_2=&&-\frac{J^2}{\Delta}\sum_{mm'll'l_1l_2}\big[\ket{G_m}\bra{G_m}\hat{T}_l\ket{E_{l_1l_2}} \nonumber \\
  &&\times\bra{E_{l_1l_2}}\hat{T}_{l'}\ket{G_{m'}}\bra{G_{m'}}\big],
\end{eqnarray}
where, $\hat{T}_{l}=\hat{S}^+_l\hat{S}^-_{l+1}+\hat{S}^-_l\hat{S}^+_{l+1}$, the summation indices $\{m,m',l,l'\}$ take values from $\{1,2,\dots,L_t\}$ and $\{l_1,l_2\}$ is summed over all states of $E_{l_1l_2}\ne E_0$.
Introducing the notations
\begin{equation}
  T_{l_1l_2}^{lm}=\bra{G_m}\hat{T}_{l}\ket{E_{l_1l_2}},
\end{equation}
\begin{equation}
  \ket{G'_{l_1l_2}}=\sum_{lm}T_{l_1l_2}^{lm}\ket{G_m},
\end{equation}
we have $\bra{G'_{l_1l_2}}=\sum_{l'm'}\bra{G_{m'}}T_{l_1l_2}^{l'm'*}$ and
\begin{eqnarray}\label{Eq_h_2_in_G_S}
  \hat{h}_2&=&-\frac{J^2}{\Delta}\sum_{mm'll'l_1l_2}\ket{G_m}T_{l_1l_2}^{lm}T_{l_1l_2}^{l'm'*}\bra{G_{m'}} \nonumber \\
  &=&-\frac{J^2}{\Delta}\sum_{l_1l_2}\ket{G'_{l_1l_2}}\bra{G'_{l_1l_2}}.
\end{eqnarray}
Using the commutation relations of the spin operators, after some algebra, we obtain
\begin{eqnarray}\label{Eq_G_l1l2_S}
  \ket{G'_{l_1l_2}}&=&\delta_{l_1,l_2-2}(\ket{G_{l_1}}+\ket{G_{l_1+1}}) \nonumber \\
  &&+\delta_{l_1-2,l_2}(\ket{G_{l_2}}+\ket{G_{l_2+1}}).
\end{eqnarray}
Inserting Eq.~\eqref{Eq_G_l1l2_S} into Eq.~\eqref{Eq_h_2_in_G_S}, we get
\begin{eqnarray}
  \hat{h}_2&=&-\frac{J^2}{\Delta}\sum_{l_1l_2}\big[\delta_{l_1,l_2-2}(\ket{G_{l_1}} +\ket{G_{l_1+1}})(\bra{G_{l_1}}+\bra{G_{l_1+1}}) \nonumber \\
  &&+\delta_{l_1-2,l_2}(\ket{G_{l_2}}+\ket{G_{l_2+1}})(\bra{G_{l_2}}+\bra{G_{l_2+1}})\big].
\end{eqnarray}
Since $\delta_{l_1,l_2-2}=1$ for $(l_1,l_2)=(m,m+2)$ with $m\in\{1,2,\dots,L_t-2\}$ and $\delta_{l_1-2,l_2}=1$ for $(l_1,l_2)=(1,L_t-1)$ or $(2,L_t)$, we have
\begin{equation}\label{Eq_h2_S}
  \hat{h}_2=-\frac{J^2}{\Delta}\sum_{m=1}^{L_t}(\ket{G_m}+\ket{G_{m+1}})(\bra{G_m}+\bra{G_{m+1}}).
\end{equation}
From Eqs.~\eqref{Eq_H_eff_def_S},~\eqref{Eq_h1_S},~and~\eqref{Eq_h2_S}, one can obtain the effective Hamiltonian
\begin{eqnarray}\label{Eq_eff_G_S}
  \hat{H}_\mathrm{eff}
  &=&-\frac{J^2}{\Delta}\sum_{m}(\ket{G_m}\bra{G_{m+1}}+\ket{G_{m+1}}\bra{G_m}) \nonumber \\
  &&+\sum_{m}(B_m+B_{m+1}+V^\mathrm{eff}_0)\ket{G_m}\bra{G_m}.
\end{eqnarray}
with the constant
$$V^\mathrm{eff}_0 =2B_0-\frac{1}{2}L_tB_0-\frac{1}{2}\sum_lB_l -\Delta\left(\frac{1}{4}L_t-1\right)-\frac{2J^2}{\Delta}.$$

In order to capture the single-quasiparticle nature of the magnon bound-states, we introduce creation operators $\hatd{b}_m$ for the magnon bound-states, which flips the spin on the $m$-th lattice site and the spin on the ($m+1$)-th lattice site from the vacuum states $\ket{\mathbf{0}}=\ket{\downarrow\downarrow\dots\downarrow}$.
That is, $\hatd{b}_m\Leftrightarrow\hat{S}^+_m\hat{S}^+_{m+1}$ and $\ket{n^{\mathrm{MBS}}_m=1} =\hatd{b}_m\ket{\mathbf{0}}\Leftrightarrow\ket{n_m=1,n_{m+1}=1} =\hat{S}^+_m\hat{S}^+_{m+1}\ket{\mathbf{0}}$.
Therefore, from Eq.~\eqref{Eq_eff_G_S}, the magnon bound-states obey the effective single-particle Hamiltonian,
\begin{equation}\label{Eq_eff_S}
  \hat{H}_\mathrm{eff}=-\frac{J^2}{\Delta}\sum_{m=1}^{L_t}(\hatd{b}_m\hat{b}_{m+1}+\mathrm{H.c.})
  +\sum_{m=1}^{L_t}\mu_m\hatd{b}_m\hat{b}_m,
\end{equation}
where $\mu_m=B_m+B_{m+1}+V^\mathrm{eff}_0$.
Since $\hat{H}_\mathrm{eff}$ commutes with $\hat{N}_\mathrm{b}=\sum_{m}\hatd{b}_m\hat{b}_m$, i.e. $[\hat{H}_\mathrm{eff},\hat{N}_\mathrm{b}]=0$, the term $V^\mathrm{eff}_0\hat{N}_\mathrm{b}$ only causes an energy shift.
Therefore, we can ignore $V^\mathrm{eff}_0$ in $\mu_m$ without changing the spectrum structure of the effective Hamiltonian~\eqref{Eq_eff_S}.

To calculate the single-particle spectrum of the effective Hamiltonian~\eqref{Eq_eff_S}, assuming the wavefunction as $\ket{\Psi}=\sum_m\psi(m)\hatd{b}_m\ket{\mathbf{0}}$, the eigenequation $\hat{H}_\mathrm{eff}\ket{\Psi}=E\ket{\Psi}$ reads as
\begin{eqnarray}\label{Eq_Harper_eff_S}
  H_\mathrm{eff}\psi(m)&=&-J^\mathrm{eff}\big[\psi(m+1)+\psi(m-1)\big]+\mu^\mathrm{eff}_m\psi(m), \nonumber \\
  &=&E\psi(m),
\end{eqnarray}
where, $J^\mathrm{eff}={J^2}/{\Delta}>0$ and $\mu^\mathrm{eff}_m=B_m+B_{m+1}=\lambda'\cos(2\pi\beta m+\delta')$ with the parameters $\lambda'=2\lambda\cos(\pi\beta)$ and $\delta'=\delta+\pi\beta$.
Obviously, the above eigenequation [Eq.~\eqref{Eq_Harper_eff_S}] reproduces the Harper equation~\cite{Lang2012} with the parameters determined by the anisotropy $\Delta$ and the frequency $\beta$.
Therefore, our effective model can be mapped onto the 2D Hofstadter model that exhibits the well-known butterfly spectrum.
Our calculations do confirm the appearance of the butterfly-like spectrum in the two-magnon bound-state band.

Moreover, the cotranslation operator $T^{(2)}$ commutes with the effective Hamiltonian, $[T^{(2)},H_\mathrm{eff}]=0$.
The action of $T^{(2)}$ on the states of the effective model reads as $T^{(2)}\psi(m)=\psi(m+q)$.
Since $\mu^\mathrm{eff}_{m+q}=\mu^\mathrm{eff}_m$, the quasi-momentum $k$ corresponding to $T^{(2)}$ [$T^{(2)}\psi(m)=e^{ikq}\psi(m)$] is the Bloch momentum of the effective model.
Hence, the quasi-momentum of the cotranslation operator is a generalization of the Bloch momentum in noninteracting 1D systems with translational symmetry to our interacting 1D systems with cotranslational symmetry, and the Chern number~\eqref{Eq_Chern_num} is an analog to the one introduced in the noninteracting 1D systems~\cite{Lang2012}.

\begin{figure}[t]
  \includegraphics[width=1.0\columnwidth]{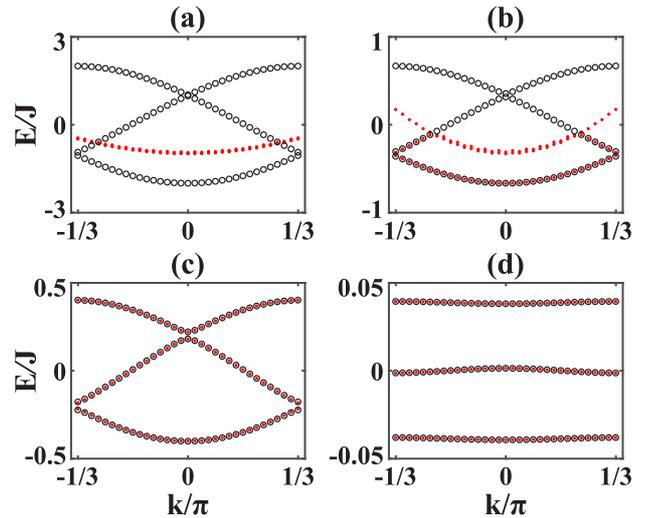}
  \caption{\label{Fig4} (Color online) The bound-state spectrum for the systems under the periodic BC.
  The black circles denote the eigenenergies for the effective Hamiltonian~\eqref{Eq_eff_S} and the red dots are the $L_t$ lowest eigenenergies for the original Hamiltonian~\eqref{Eq_XXZ_Ham}.
  The parameters are chosen as $\lambda/J=0.04$, $\beta=1/3$, $\delta=\pi/6$, $L_t=99$, and different values of $\Delta/J$:
  (a) $\Delta/J=1$,
  (b) $\Delta/J=3$,
  (c) $\Delta/J=5$,
  (d) $\Delta/J=100$.
  The eigenenergies $E/J$ are added with a constant $\Delta/J+2J/\Delta$.}
\end{figure}

Denoting $\psi_{n,k}=\psi_{n,k}(m)$ as the common eigenstates of $H_\mathrm{eff}$ and $T^{(2)}$, that is $H_\mathrm{eff}\psi_{n,k}=E_n\psi_{n,k}$ and $T^{(2)}\psi_{n,k}=e^{ikq}\psi_{n,k}$,
we have $\psi_{n,k}(m)=e^{ikm}\phi_{n,k}(m)$ with $\phi_{n,k}(m+q)=\phi_{n,k}(m)$ due to the Bloch's theorem.
In terms of $\phi_{n,k}$, the eigenequation~\eqref{Eq_Harper_eff_S} can be expressed as
\begin{eqnarray}
  E\phi_{n,k}(m)&=&-J^\mathrm{eff}\big[e^{ik}\phi_{n,k}(m+1)+e^{-ik}\phi_{n,k}(m-1)\big] \nonumber \\
  &&+\mu^\mathrm{eff}_m\phi_{n,k}(m)
\end{eqnarray}
with $q$ independent functions $\phi_{n,k}(m)$.
Then the single-particle spectrum of the effective model is given by diagonalizing the $q\times q$ matrix for the effective Hamiltonian.
In Fig.~\ref{Fig4}, we show the spectrum of the effective Hamiltonian~\eqref{Eq_eff_S} and the magnon bound-state bands of the original Hamiltonian~\eqref{Eq_XXZ_Ham}.
Under sufficiently strong interactions, the effective model~\eqref{Eq_eff_S} well describes the magnon bound-states in the original system~\eqref{Eq_XXZ_Ham}.

Under the periodic BC, the Bloch momentum $k$ for the effective Hamiltonian~\eqref{Eq_eff_S} is a good quantum number and $(k,\delta)$ makes up a 2D parameter space to define the Chern number.
We calculate the Chern numbers for the whole spectrum of the effective Hamiltonian using the same parameters given in Tab.~\ref{Tab-TwoMagnonChernNumber}.
The results are consistent with the ones in Tab.~\ref{Tab-TwoMagnonChernNumber} for the original system~\eqref{Eq_XXZ_Ham}.
This means that the topological features of the two-magnon bound-states in the original Hamiltonian~\eqref{Eq_XXZ_Ham} can be described by the effective single-particle Hamiltonian~\eqref{Eq_eff_S}.

So far, we only consider the system under the periodic BC.
However, the BCs may have strong effects on the spectrum and the emergence of the edge states.
Below we consider the system under the open BC and its effective single-particle Hamiltonian.

\begin{figure}[t]
  \includegraphics[width=1.0\columnwidth]{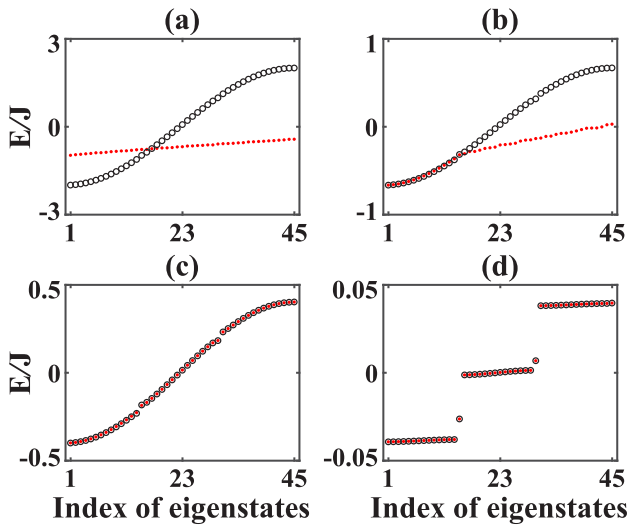}
  \caption{\label{Fig5} (Color online) The bound-state spectrum for the systems under the open BC.
  The black circles denote the eigenenergies for the effective Hamiltonian~\eqref{Eq_eff_OBC_S} and the red dots are the $L_t$ lowest eigenenergies for the original Hamiltonian~\eqref{Eq_XXZ_Ham}.
  The parameters are given as $\lambda/J=0.04$, $\beta=1/3$, $\delta=\pi/6$, $L_t=46$, and different values of $\Delta/J$:
  (a) $\Delta/J=1$,
  (b) $\Delta/J=3$,
  (c) $\Delta/J=5$,
  (d) $\Delta/J=100$.
  The energies $E/J$ are added with a constant $\Delta/J+2J/\Delta$.}
\end{figure}

Under the open BC, the dimension of the Hilbert space spanned by the ground states $\ket{G_l}=\ket{l,l+1}$ is $L_t-1$ rather than $L_t$ for the system under the periodic BC.
The Hilbert space for the magnon bound-states is then $\mathcal{U}_1=\{\ket{G_l}:1\le l\le L_t-1\}$, and the corresponding projection operator onto $\mathcal{U}_1$ is still Eq.~\eqref{Eq_Project_U0_S} but with the summation index $l$ from $1$ to $L_t-1$.
Following the similar procedure of deriving the effective Hamiltonian under the periodic BC,
we obtain the effective Hamiltonian $\left.\hat{H}_\mathrm{eff}^\mathrm{O} =\hat{h}_0^\mathrm{O}+\hat{h}_1^\mathrm{O}+\hat{h}_2^\mathrm{O}\right.$ with
\begin{equation}
  \hat{h}_1^\mathrm{O}=\sum_{m=1}^{L_t-1}\left(B_m+B_{m+1}-\tfrac{1}{2}\sum_lB_l\right)\ket{G_m}\bra{G_m},
\end{equation}
and
\begin{equation}
  \hat{h}_2^\mathrm{O}=-\frac{J^2}{\Delta}\sum_{m=1}^{L_t-2}(\ket{G_m} +\ket{G_{m+1}})(\bra{G_m}+\bra{G_{m+1}}).
\end{equation}
Introducing the creation operators $\hatd{b}_m$ (in which the index $m=\{1,2,\cdots,L_t-1\}$),
the effective Hamiltonian reads as
\begin{eqnarray}\label{Eq_eff_OBC_S}
  \hat{H}_\mathrm{eff}^\mathrm{O}&=& -\frac{J^2}{\Delta}\sum_{m=1}^{L_t-2}(\hatd{b}_m\hat{b}_{m+1}+\mathrm{H.c.})
  +\sum_{m=1}^{L_t-1}\mu_m\hatd{b}_m\hat{b}_m \nonumber \\
  &&+\frac{J^2}{\Delta}(\hatd{b}_1\hat{b}_1+\hatd{b}_{L_t-1}\hat{b}_{L_t-1}),
\end{eqnarray}
with $\mu_m=B_m+B_{m+1}+V^\mathrm{eff}_0$.
The validity of the effective Hamiltonian under the open BC is shown in Fig.~\ref{Fig5}.

\section{Topological equivalence between the effective Hamiltonian and the original Hamiltonian\label{SecAppC}}

\begin{figure}[t]
  \includegraphics[width=1.0\columnwidth]{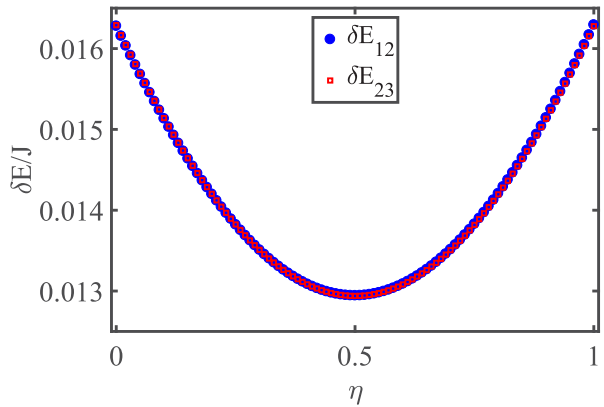}
  \caption{\label{Fig6} (Color online) The energy gaps versus the deforming parameter $\eta$. We consider the system \eqref{Eq_H_deform} under the periodic BC. The parameters are chosen as $\Delta/J=100$, $\lambda/J=0.04$, $\beta=1/3$, and $L_t=99$.}
\end{figure}

Since the mapping from the original model to the effective model is in the context of the second-order perturbation theory, one can not guarantee that such a perturbation treatment retains the topological nature of the original model.
Below, through introducing an auxiliary model, we demonstrate the topological equivalence between the original model and the effective model numerically.

According to the topological band theory~\cite{KaneRev2010, QiRev2011}, the gap in a topological system is associated with an index that characterizes the topological nature of the bands below this gap.
In fact, such an index is a topological invariant and it equals to the sum of the Chern numbers of the bands below this gap.
Two gapped systems belong to the same topological state if they can be continuously deformed from one into the other without gap closing.
And vice versa, when a system with a topological index is continuously deformed into another system with different topological index (i.e. the two systems are not topologically equivalent), the bands will invert and the gap closes during the deforming process.
Correspondingly, a so-called topological phase transition occurs in the gap closing process.

To show the topological equivalence between the original model and the effective model,
we introduce an auxiliary model,
\begin{equation}\label{Eq_H_deform}
  \hat{H}^A(\eta)=\hat{H}+\eta(\hat{H}_\mathrm{eff}-\hat{H}),
\end{equation}
which is a combination of the original model $\hat{H}$ described by the Hamiltonian~\eqref{Eq_XXZ_Ham} and the effective model $\hat{H}_\mathrm{eff}$ described by the Hamiltonian~\eqref{Eq_eff_S}.
Here, $\eta$ is the deforming parameter.
To discuss the two-magnon excitations, $\hat{H}$ and $\hat{H}_\mathrm{eff}$ are both restricted in the two-magnon Hilbert space.
Therefore $\hat{H}_\mathrm{eff}$ can be explicitly rewritten in the form of
\begin{eqnarray}
  \hat{H}_\mathrm{eff}=&&-\frac{J^2}{\Delta}\sum_m(\hat{S}^+_m\hat{S}^+_{m+1}\hat{S}^-_{m+2}\hat{S}^-_{m+1}+\mathrm{H.c.}) \nonumber \\
  &&+\sum_{m}\mu_m\hat{S}^+_m\hat{S}^+_{m+1}\hat{S}^-_{m+1}\hat{S}^-_{m},
\end{eqnarray}
with $\mu_m=B_m+B_{m+1}+V^\mathrm{eff}_0$.

Obviously, $\hat{H}^A=\hat{H}$ for $\eta=0$ and $\hat{H}^A=\hat{H}_\mathrm{eff}$ for $\eta=1$.
Thus, if the parameter $\eta$ is continuously changed from $0$ to $1$, the system $\hat{H}^A$ continuously deforms from $\hat{H}$ into $\hat{H}_\mathrm{eff}$.
The energy gap between the $j$-th subband and the $(j+1)$-th subband for the magnon bound-state band of $\hat{H}^A$ can be given as
\begin{equation}
  \delta E_{j,j+1}=\min_{k,\delta}\big[E_{j+1}(k,\delta)-E_j(k,\delta)\big].
\end{equation}
To explore whether the effective model $H$ and the original model $\hat{H}_\mathrm{eff}$ are topological equivalent, we calculate the energy gap versus the deforming parameter $\eta$.
As an example, for $\beta=1/3$, there are three subbands in the magnon bound-state band and therefore there are two energy gaps $\delta E_{12}$ and $\delta E_{23}$.
As shown in Fig.~\ref{Fig6}, the energy gaps are nonzero for all $\eta$, that is the energy gaps keep open during the whole deforming process.
This indicates that the original model and the effective model are topological equivalent.



\begin{thebibliography}{99}

\bibitem{Bethe1931}
  H. Bethe,
  Zur Theorie der Metalle,
  Z. Phys. \textbf{71}, 205 (1931).

\bibitem{Wortis1963}
  M. Wortis,
  Bound States of Two Spin Waves in the Heisenberg Ferromagnet,
  Phys. Rev. \textbf{132}, 85 (1963).

\bibitem{Hanus1963}
  J. Hanus,
  Bound States in the Heisenberg Ferromagnet,
  Phys. Rev. Lett. \textbf{11}, 336 (1963).

\bibitem{BlochNatPhys2013}
  T. Fukuhara, A. Kantian, M. Endres, M. Cheneau, P. Schau\ss, S. Hild, D. Bellem, U. Schollw\"{o}ck, T. Giamarchi, C. Gross, I. Bloch, and S. Kuhr,
  Quantum dynamics of a mobile spin impurity,
  Nat. Phys. \textbf{9}, 235 (2013).

\bibitem{BlochNature2013}
  T. Fukuhara, P. Schau\ss, M. Endres, S. Hild, M. Cheneau, I. Bloch, and C. Gross,
  Microscopic observation of magnon bound states and their dynamics,
  Nature \textbf{502}, 76 (2013).

\bibitem{KaneRev2010}
  M. Z. Hasan and C. L. Kane,
  \textit{Colloquium}: Topological insulators,
  Rev. Mod. Phys. \textbf{82}, 3045 (2010).

\bibitem{QiRev2011}
  X.-L. Qi and S.-C. Zhang,
  Topological insulators and superconductors,
  Rev. Mod. Phys. \textbf{83}, 1057 (2011).

\bibitem{MooreNature2010}
  J. E. Moore,
  The birth of topological insulators,
  Nature \textbf{464}, 194 (2010).

\bibitem{Schnyder2008}
  A. P. Schnyder, S. Ryu, A. Furusaki, and A. W. W. Ludwig,
  Classification of topological insulators and superconductors in three spatial dimensions,
  Phys. Rev. B \textbf{78}, 195125 (2008).

\bibitem{Kitaev2009}
  A. Kitaev,
  Periodic table for topological insulators and superconductors,
  AIP Conf. Proc. \textbf{1134}, 22 (2009).

\bibitem{MH2010}
  H. Katsura, N. Nagaosa, and P. A. Lee,
  Theory of the Thermal Hall Effect in Quantum Magnets,
  Phys. Rev. Lett. \textbf{104}, 066403 (2010).

\bibitem{ExpMH2010}
  Y. Onose, T. Ideue, H. Katsura, Y. Shiomi, N. Nagaosa, and Y. Tokura,
  Observation of the Magnon Hall Effect,
  Science \textbf{329}, 297 (2010).

\bibitem{Li2013}
  L. Zhang, J. Ren, J.-S. Wang, and B. Li,
  Topological magnon insulator in insulating ferromagnet,
  Phys. Rev. B \textbf{87}, 144101 (2013).

\bibitem{Pereiro2014}
  M. Pereiro, D. Yudin, J. Chico, C. Etz, O. Eriksson, and A. Bergman,
  Topological excitations in a kagome magnet,
  Nat. Commun. \textbf{5}, 4815 (2014).

\bibitem{Sachdev2014}
  M. Punk,	D. Chowdhury, and S. Sachdev,
  Topological excitations and the dynamic structure factor of spin liquids on the kagome lattice,
  Nat. Phys. \textbf{10}, 289 (2014).

\bibitem{Ong2015}
  M. Hirschberger, R. Chisnell, Y. S. Lee, and N. P. Ong,
  Thermal Hall Effect of Spin Excitations in a Kagome Magnet,
  Phys. Rev. Lett. \textbf{115}, 106603 (2015).

\bibitem{Lee2015}
  R. Chisnell, J. S. Helton, D. E. Freedman, D. K. Singh, R. I. Bewley, D. G. Nocera, and Y. S. Lee,
  Topological Magnon Bands in a Kagome Lattice Ferromagnet,
  Phys. Rev. Lett. \textbf{115}, 147201 (2015).

\bibitem{FTI2015}
  J. Maciejko and G. A. Fiete,
  Fractionalized topological insulators,
  Nat. Phys. \textbf{11}, 385 (2015).

\bibitem{Grusdt2015}
  F. Grusdt, N. Y. Yao, D. Abanin, M. Fleischhauer, and E. Demler,
  Interferometric measurements of many-body topological invariants using mobile impurities,
  Nat. Commun. \textbf{7}, 11994 (2016).

\bibitem{Chen2012}
  X. Chen, Z.-C. Gu, Z.-X. Liu, and X.-G. Wen,
  Symmetry-Protected Topological Orders in Interacting Bosonic Systems,
  Science \textbf{338}, 1604 (2012).

\bibitem{Wang2014}
  C. Wang, A. C. Potter, and T. Senthil,
  Classification of Interacting Electronic Topological Insulators in Three Dimensions,
  Science \textbf{343}, 629 (2014).

\bibitem{TKNN}
  D. J. Thouless, M. Kohmoto, M. P. Nightingale, and M. den Nijs,
  Quantized Hall Conductance in a Two-Dimensional Periodic Potential,
  Phys. Rev. Lett. \textbf{49}, 405 (1982).

\bibitem{Niu1985}
  Q. Niu, D. J. Thouless, and Y.-S. Wu,
  Quantized Hall conductance as a topological invariant,
  Phys. Rev. B \textbf{31}, 3372 (1985).

\bibitem{WangPRL2010}
  Z. Wang, X.-L. Qi, and S.-C. Zhang,
  Topological Order Parameters for Interacting Topological Insulators,
  Phys. Rev. Lett. \textbf{105}, 256803 (2010).

\bibitem{WangPRX2012}
  Z. Wang and S.-C. Zhang,
  Simplified Topological Invariants for Interacting Insulators,
  Phys. Rev. X \textbf{2}, 031008 (2012).

\bibitem{iontrapEXP1}
  P. Richerme, Z.-X. Gong, A. Lee, C. Senko, J. Smith, M. Foss-Feig, S. Michalakis, A. V. Gorshkov, and C. Monroe,
  Non-local propagation of correlations in quantum systems with long-range interactions,
  Nature \textbf{511}, 198 (2014).

\bibitem{iontrapEXP2}
  P. Jurcevic, B. P. Lanyon, P. Hauke, C. Hempel, P. Zoller, R. Blatt, and C. F. Roos,
  Quasiparticle engineering and entanglement propagation in a quantum many-body system,
  Nature \textbf{511}, 202 (2014).

\bibitem{Grass2015}
  T. Gra\ss, C. Muschik, A. Celi, R. W. Chhajlany, and M. Lewenstein,
  Synthetic magnetic fluxes and topological order in one-dimensional spin systems,
  Phys. Rev. A \textbf{91}, 063612 (2015).

\bibitem{Lang2012}
  L.-J. Lang, X. Cai, and S. Chen,
  Edge States and Topological Phases in One-Dimensional Optical Superlattices,
  Phys. Rev. Lett. \textbf{108}, 220401 (2012).

\bibitem{Kannappan2009}
  Pl. Kannappan,
  in \textit{Functional Equations and Inequalities with Applications, Springer Monographs in Mathematics}
  (Springer-Verlag US, 2009).

\bibitem{Mei2012}
  F. Mei, S.-L. Zhu, Z.-M. Zhang, C. H. Oh, and N. Goldman,
  Simulating ${Z}_{2}$ topological insulators with cold atoms in a one-dimensional optical lattice,
  Phys. Rev. A \textbf{85}, 013638 (2012).

\bibitem{Zhu2013}
  S.-L. Zhu, Z.-D. Wang, Y.-H. Chan, and L.-M. Duan,
  Topological Bose-Mott Insulators in a One-Dimensional Optical Superlattice,
  Phys. Rev. Lett. \textbf{110}, 075303 (2013).

\bibitem{Mei2014}
  F. Mei, J.-B. You, D.-W. Zhang, X. C. Yang, R. Fazio, S.-L. Zhu, and L. C. Kwek,
  Topological insulator and particle pumping in a one-dimensional shaken optical lattice,
  Phys. Rev. A \textbf{90}, 063638 (2014).

\bibitem{Qin2014}
  X. Qin, Y. Ke, X. Guan, Z. Li, N. Andrei, and C. Lee,
  Statistics-dependent quantum co-walking of two particles in one-dimensional lattices with nearest-neighbor interactions,
  Phys. Rev. A \textbf{90}, 062301 (2014).

\bibitem{Harper1955}
  P. G. Harper,
  Single Band Motion of Conduction Electrons in a Uniform Magnetic Field,
  Proc. Phys. Soc. A \textbf{68}, 874 (1955).

\bibitem{Hofstadter1976}
  D. R. Hofstadter,
  Energy levels and wave functions of Bloch electrons in rational and irrational magnetic fields,
  Phys. Rev. B \textbf{14}, 2239 (1976).

\bibitem{Hatsugai1993a}
  Y. Hatsugai,
  Edge states in the integer quantum Hall effect and the Riemann surface of the Bloch function,
  Phys. Rev. B \textbf{48}, 11851 (1993).

\bibitem{Hatsugai1993b}
  Y. Hatsugai,
  Chern number and edge states in the integer quantum Hall effect,
  Phys. Rev. Lett. \textbf{71}, 3697 (1993).

\bibitem{Mandel2003}
  O. Mandel, M. Greiner, A. Widera, T. Rom, T. W. H\"{a}nsch, and I. Bloch,
  Coherent Transport of Neutral Atoms in Spin-Dependent Optical Lattice Potentials,
  Phys. Rev. Lett. \textbf{91}, 010407 (2003).

\bibitem{Widera2004}
  A. Widera, O. Mandel, M. Greiner, S. Kreim, T. W. H\"{a}nsch, and I. Bloch,
  Entanglement Interferometry for Precision Measurement of Atomic Scattering Properties,
  Phys. Rev. Lett. \textbf{92}, 160406 (2004).

\bibitem{Gross2010}
  C. Gross, T. Zibold, E. Nicklas, J. Est\`{e}ve, and M. K. Oberthaler,
  Nonlinear atom interferometer surpasses classical precision limit,
  Nature \textbf{464}, 1165 (2010).

\bibitem{Qin2016}
  X. Qin, F. Mei, Y. Ke, L. Zhang, and C. Lee,
  Topological magnon bound-states in quantum Heisenberg chains,
  arXiv: 1602.03217.

\bibitem{DiLiberto2016}
  M. Di Liberto, A. Recati, I. Carusotto, and C. Menotti,
  Two-body physics in the Su-Schrieffer-Heeger model,
  Phys. Rev. A \textbf{94}, 062704 (2016).

\bibitem{Marques2017}
  A. M. Marques and R. G. Dias,
  Multihole edge states in Su-Schrieffer-Heeger chains with interactions,
  Phys. Rev. B \textbf{95}, 115443 (2017).

\bibitem{Resta2000}
  R. Resta,
  Manifestations of Berry's phase in molecules and condensed matter,
  J. Phys.: Condens. Matter \textbf{12}, R107 (2000).

\bibitem{Fukui2005}
  T. Fukui, Y. Hatsugai, and H. Suzuki,
  Chern Numbers in Discretized Brillouin Zone: Efficient Method of Computing (Spin) Hall Conductances,
  J. Phys. Soc. Jpn. \textbf{74}, 1674 (2005).

\bibitem{Takahashi1977}
  M. Takahashi,
  Half-filled Hubbard model at low temperature,
  J. Phys. C: Solid State Phys. \textbf{10}, 1289 (1977).

\end{thebibliography}
\end{document}